\documentclass[final,1p,times,twocolumn]{elsarticle}

\usepackage{amssymb}

\journal{Physics Letters B}

\begin{document}

\begin{frontmatter}



\title{Supersymmetric Nonlinear Sigma Model in AdS$_5$}


\author{Jonathan Bagger and Jingsheng Li}
\ead{bagger@jhu.edu, jsli@pha.jhu.edu}

\address{Department of Physics and Astronomy, Johns Hopkins University, 3400 North Charles Street, Baltimore, MD 21218}

\begin{abstract}
We construct the supersymmetric nonlinear sigma model in a fixed AdS$_5$ background.  We use component fields and find that the complex bosons must be the coordinates of a hyper-K\"ahler manifold that admits a Killing vector satisfying an inhomogeneous tri-holomorphic condition.  We propose boundary conditions that map the on-shell bulk hypermultiplets into off-shell chiral multiplets on 3-branes that foliate the bulk.  The supersymmetric AdS$_5$ isometries reduce to superconformal transformations on the brane fields.
\end{abstract}




\end{frontmatter}

\newcommand{\nn}{\nonumber}

\section{Introduction}

Over thirty years ago, Zumino revealed the relation between supersymmetry and complex\break geometry \cite{Zumino}:  In four flat dimensions, the target space of an ${\cal N} =1$ supersymmetric nonlinear sigma model must be K\"ahler.  Soon thereafter, Alvarez-Gaum\'e and Freedman \cite{Freedman} found that ${\cal N} =2$ supersymmetry restricts the target space to be hyper-K\"ahler (see also \cite{Curtright}).  In five and six flat dimensions, nonlinear sigma models also require hyper-K\"ahler geometry \cite{Townsend}.  In each case, the model is specified by a real K\"ahler potential $K$, and for hyper-K\"ahler manifolds, a holomorphic covariantly constant anti-symmetric tensor $\Omega^{ij}$.  Mass terms and Yukawa couplings are determined by a holomorphic superpotential $P$.

Given the explosion of interest in AdS/CFT, one would also like to know the most general\break supersymmetric coupling of hypermultiplets in AdS backgrounds, and in particular, AdS$_5$.  During the the past few years, this issue was examined in ${\cal N} = 2$ superspace \cite{Kuzenko}, as well as in warped ${\cal N} = 1$ superspace \cite{BX}.  In this note we eschew the superspace formalism and work directly in terms of component fields.  We construct the most general Lagrangian and transformations laws, and provide independent confirmation of the results of \cite{BX}.  We also find boundary conditions that reduce the on-shell hypermultiplets to off-shell chiral multiplets on flat 3-branes embedded in AdS$_5$.

The letter is organized as follows.  In sect.\ 2 we set the notation by reviewing the most general hypermultiplet coupling in flat five-dimensional spacetime.  In sect.\ 3 we construct the most general hypermultiplet coupling in AdS$_5$.  We close the algebra, find the action, and determine all constraints on the theory.  We find the target space must be a hyper-K\"ahler manifold endowed with a holomorphic Killing vector that obeys a particular inhomogeneous tri-holomorphic condition.  In sect.\ 4 we propose boundary conditions that transform the on-shell bulk hypermultiplets into off-shell brane chiral multiplets.  With our boundary conditions, the super AdS$_5$ isometries become superconformal transformations on the branes.  We conclude with a summary in sect.\ 5.

\section{Hypermultiplet in Five Flat Dimensions}

We start by reviewing supersymmetric nonlinear sigma model in five flat dimensions, where supersymmetry algebra is:
\begin{equation}
\label{alg}
\{\mathbb{Q},\bar{{\mathbb{Q}}}\}=2\gamma^MP_M+2Z .
\end{equation}
We use four-component Dirac formalism and define $\bar{\mathbb{Q}}\equiv\mathbb{Q}^{\dag}\gamma^{0}$. Here $Z$ is a real central charge that commutes with other generators.

A set of $n$ five-dimensional hypermultiplet contains $2n$ complex scalars and $n$ Dirac fermions.  The $2n$ complex scalars $A^i$ parametrize a hyper-K\"ahler manifold, endowed with a covariantly constant K\"ahler metric $g_{ij^*}$ and a holomorphic covariantly constant anti-symmetric tensor $\Omega^{ij}$, with ${\Omega^i}_{k^*}{\bar\Omega}^{k^*}_{\ j}=-\delta^i{}_j$, where ${\Omega^i}{}_{k^*}\equiv\Omega^{ij}g_{jk^*}$ and ${\bar\Omega}^{k^*}_{\ j}\equiv{({\Omega}^k{}_{j^*})}^*$.  We combine the $n$ Dirac spinors into $2n$ symplectic Majorana spinors, $\Psi^i=(\chi^i, \Omega^i{}_{j^*}\bar{\chi}^{j^*})^T$, where $i=1,...,2n$.

In flat space, the supersymmetry transformations are given by
\begin{eqnarray}\label{5DSUSY}
\delta A^i&=&\sqrt{2}\bar{\epsilon}_{+}\Psi^i\nn\\
\delta\Psi^i&=&\sqrt{2}(i\gamma^M\epsilon_{+}\partial_MA^i+i\Omega^i{}_{j^*}\gamma^M\epsilon_{-}\partial_MA^{*j^*}
+iX^i\epsilon_{+}+i\Omega^i{}_{j^*}\bar{X}^{j^*}\epsilon_{-})-\Gamma^i_{jk}\delta A^j\Psi^k,
\end{eqnarray}
where $\Gamma^i_{jk}$ is the Christoffel symbol of the hyper-K\"ahler manifold, $X^i$ is a tri-holomorphic Killing vector,
\begin{eqnarray}\label{tri-holo condition}
\nabla_iX_{j^*}+\nabla_{j^*}\bar{X}_i&=&0
 \nn\\
\nabla_jX^{i}+{\bar\Omega}^{k^*}_{\ j}\nabla_{k^*}\bar{X}^{l^*}\Omega^i{}_{l^*}&=&0,
\end{eqnarray}
and the supersymmetry parameters $\epsilon_\pm$ are constant spinors:
\begin{equation}
\epsilon_+= \left(\begin{array}{c}
-\eta\\
\bar{{\epsilon}}\end{array}\right),\qquad
\epsilon_-= \left(\begin{array}{c}
\epsilon\\
\bar{{\eta}}\end{array}\right).
\end{equation}
The supersymmetry algebra closes with the help of the fermion equations of motion:
\begin{equation}
\label{fermieom}
i\gamma^M {\cal D}_M\Psi^i-i\nabla_j X^i\Psi^j+\frac{1}{2}R^i{}_{jk^*l}(\bar{\Psi}^{k^*}\Psi^j)\Psi^l=0.
\end{equation}
where $\mathcal{D}_M\Psi^i=\partial_M\Psi^i + \Gamma^{i}_{jk} \partial_m A^j \Psi^k$.  Consistency requires that $X^i$ be a tri-holomorphic Killing vector (\ref{tri-holo condition}) on the target space.

It is useful to re-write the algebra (\ref{alg}) in two-component notation \cite{Wess-Bagger}.  The supersymmetry generator $\mathbb{Q}$ splits naturally into two Weyl spinors, $\mathbb{Q}=(Q, \bar{S})^T$.  In this notation, the algebra takes a form similar to $\mathcal{N}=2$ in four dimensions:
\begin{eqnarray}
  \{Q_{\alpha}, \bar{Q}_{\dot{\beta}}\}\ =\  \{S_{\alpha}, \bar{S}_{\dot{\beta}}\} &=& 2\sigma_{\alpha\dot{\beta}}^mP_m \label{5d Q1Q1commutator}\nn\\
  \{Q_{\alpha}, S_{\beta}\} &=& 2\epsilon_{\alpha\beta}\mathcal{Z}. \label{5d Q1Q2commutator}
  \end{eqnarray}
The real central charge $Z$ combines with $P_5$ form a complex central charge ${\cal Z}=Z-iP_5$ in four dimensions.

The $2n$ complex bosons $A^i$ and $2n$ Weyl fermions $\chi^i$ have the following supersymmetry transformations:
\begin{eqnarray}
\delta A^i & = & \sqrt{{2}}(\epsilon\chi^i+\Omega^i{}_{j^*}\bar{\eta}\bar{\chi}^{j^*})\nn\\
\delta\chi^i & = & \sqrt{{2}}(i\sigma^m\bar{\epsilon}\partial_mA^i-i\Omega^i{}_{j^*}\sigma^m\bar{\eta}\partial_mA^{*j}
-\Omega_{\ j^*}^i\partial_{5}A^{j^*}\epsilon
+\partial_{5}A^i\eta
-i\Omega_{\ j^*}^i\bar{X}^{j^*}\epsilon
-iX^i\eta)\nn\\
&&\ -\ \Gamma^i_{jk}\delta A^j\psi^k .
\label{N=2_SUSY}
\end{eqnarray}
Imposing (\ref{tri-holo condition}) and using the fermion equations of motion, one finds that the first and second supersymmetry transformations close into a diffeomorphism,
\begin{eqnarray}\label{N=2 analytic diffeo}
\delta_{X} A^i &= & \xi X^i \nn \\
\delta_{X} \chi^i & = & \xi X^i_j\chi^j,
\end{eqnarray}
where the parameter $\xi=2i(\epsilon\eta-\bar{\epsilon}\bar{\eta})$.  The diffeomorphism (\ref{N=2 analytic diffeo}) is an isometry of the target space.  It leaves the metric and the anti-symmetric tensor invariant, $\delta_Xg_{ij^*}=\delta_X\Omega^{ij}=0$.

Given the equations of motion, it is not hard to work backwards to determine the invariant action.  We find
\begin{eqnarray}\label{action}
S&=& \int dx^5 e\, \Big\{-g_{ij^*}\partial^MA^i\partial_MA^{*j^*} - \frac{i}{2} g_{ij^*}\bar{\Psi}^{j^*}\gamma^M\mathcal{D}_M\Psi^i
- {\cal V} \nonumber\\
&&\ +\ \frac{i}{4}g_{ij^*}\nabla_kX^i\bar{\Psi}^{j^*}\Psi^k-\frac{i}{4}g_{ij^*}\nabla_{k^*}\bar{X}^{j^*}\bar{\Psi}^{k^*}\Psi^i
+\frac{1}{8}R_{ij^*kl^*}(\bar{\Psi}^{j^*}\Psi^i)(\bar{\Psi}^{l^*}\Psi^k)\Big\}.
\end{eqnarray}
This is the action for the supersymmetric nonlinear sigma model in five flat dimensions.  The target space is a hyper-K\"ahler manifold; the Killing vector $X^i$ determines the potential,
\begin{equation}
{\cal V} =  g_{ij^*}X^i \bar{X}^{j^*}.
\end{equation}

\section{Hypermultiplets in AdS$_{5}$}

We are now in position to discuss sigma models on AdS$_5$.  We choose a coordinate system in which AdS$_5$ metric is $ds^{2}=e^{-2kz}\eta_{mn}dx^mdx^n+dz^{2}$, where $z=x^5$.  There are 15 bosonic isometries in this space:  $(P_a, M_{ab}, D, K_a)$.  The names of isometries are chosen to highlight their one-to-one correspondence with the generators of the four-dimensional conformal group.

The supergroup of AdS$_5$ is called SU(2,2$|$1).  Its bosonic sector contains the 15 AdS$_5$ isometries plus an extra U(1) symmetry, the lift of the four-dimensional superconformal $R$-symmetry.  The SU(2,2$|$1) commutation relations are as follows,
\begin{eqnarray}\label{algads}
\{Q_\alpha, \bar{{Q}}_{\dot\alpha}\}&=& 2\sigma^a_{\alpha\dot\alpha} P_a\nn\\
\{S_\alpha, \bar{{S}}_{\dot\alpha}\}&=&2\sigma^a_{\alpha\dot\alpha} K_a\nn\\
\{Q_\alpha, S_\beta\}&=&4iM_{\alpha\beta}+2i\epsilon_{\alpha\beta} D-6\epsilon_{\alpha\beta} U.
\end{eqnarray}
In any curved space, the supersymmetry transformation parameter must obey the Killing spinor equation,
\begin{equation}
\mathcal{D}_M\epsilon_\pm=\partial_M\epsilon_\pm+\frac{1}{4}\omega_M^{AB}\gamma_{AB}\epsilon_\pm=\pm\frac{i}{2}k\gamma_M\epsilon_\pm.
\end{equation}
For the case at hand, it has two independent solutions, specified by the constant spinors $\epsilon$ and $\eta$:
\begin{equation}
\epsilon_{+}=\left(\begin{array}{c}
-e^{\frac{1}{2}kz}\eta\\
e^{-\frac{1}{2}kz}\bar{\epsilon}-ike^{-\frac{1}{2}kz}x^m\delta_m^{a}\bar{\sigma}_{a}\eta\end{array}\right).
\end{equation}
The parameter $\epsilon_{-}$ is the symplectic dual of $\epsilon_{+}$:
\begin{equation}
\epsilon_{-}\equiv\left(\begin{array}{c}
e^{-\frac{1}{2}kz}\epsilon-ike^{-\frac{1}{2}kz}x^m\delta_m^{a}\sigma_{a}\bar{\eta}\\
e^{\frac{1}{2}kz}\bar{\eta}\end{array}\right).
\end{equation}

As in flat space, the $n$ AdS$_5$ hypermultiplets contain $2n$ complex scalars $A^i$ and $2n$ Weyl fermions $\chi^i$.  Where appropriate, we collect the fermions into $2n$ symplectic Majorana spinors $\Psi^i=(\chi^i, \Omega^i{}_{j^*}\bar{\chi}^{j^*})^T$ that obey the constraint $\bar{\epsilon}_{+}\Psi^i=-\Omega^i{}_{j^*}\bar{\Psi}^{j^*}\epsilon_{-}$.

To find the transformations, we write down the most general expressions based on five-dimensional Lorentz covariance, target space diffeomorphism covariance, and the requirement that every slice $z=c$ have ${\cal N}=1$ supersymmetry.  That is enough to restrict the transformations to be precisely of the form (\ref{5DSUSY}), where the target-space manifold is K\"ahler.  Closure on the bosons tells us that $\Omega^{ij}$ must be holomorphic and covariantly constant, so the target-space manifold is also hyper-K\"ahler.  Closure on the fermions implies that $X^i$ is holomorphic and that it satisfies the following constraint:
\begin{equation}\label{Modifiedtri-holo}
\nabla_j X^i+\Omega^i{}_{j^*}\nabla_{k^*}{\bar X}^{j^*}{\bar\Omega}^{k^*}{}_{j}=-3ik\delta^i{}_j.
\end{equation}
This result differs from the tri-holomorphic condition (\ref{tri-holo condition}) by the nonzero imaginary piece on the right-hand side. It is the same condition that was found in \cite{BX} using superspace techniques.  It is similar to the condition found in \cite{BK} for ${\cal N}=2$ hyper-K\"ahler models in AdS$_4$.

The fermion equations of motion also follow from the closure,
\begin{equation}\label{AdS Dirac}
i\bar{\sigma}^m \mathcal{D}_m\chi^i+\Omega^i{}_{j^*}\mathcal{D}_5\bar{\chi}^{j^*}-\frac{k}{2}\Omega^i{}_{j^*}\bar{\chi}^{j^*}+i\Omega^i{}_{j^*}\nabla_{k^*}\bar{X}^{j^*}\bar{\chi}^{k^*}-\frac{1}{2}g^{im^*}R_{jk^*lm^*}(\chi^j\chi^l)\bar{\chi}^{k^*}=0.
\end{equation}
The Killing condition,
\begin{equation}\label{AdSKillingCondition}
\nabla_iX_{j^*}+\nabla_{j^*}\bar{X}_i=0,
\end{equation}
follows from requiring that $\delta_{\epsilon}$ and $\delta_{\eta}$, acting on (\ref{AdS Dirac}), produce the same bosonic equations of motion.  Therefore $X^i$ must be a Killing vector that satisfies the inhomogeneous tri-holomorphic condition (\ref{Modifiedtri-holo}) on the hyper-K\"ahler manifold.

In accord with the algebra (\ref{algads}), the anti-commutator of $\{Q,S\}$ generates a $U$ transformation,
\begin{eqnarray}\label{U(1)}
\delta_U A^i&=&\xi X^i\nonumber\\
\delta_U \chi^i&=&\xi X^i_j\chi^j+\frac{3}{2}ik\xi\chi^i,
\end{eqnarray}
where $\xi=2i\left(\epsilon\eta-\bar{\epsilon}\bar{\eta}\right)$.  This is an isometry of the hyper-K\"ahler manifold, and $\delta_U g_{ij^*}=0$, as required.  It is perhaps more interesting to note that $\delta_U \Omega^{ij}=3ik\xi \Omega^{ij}$.  The isometry rotates the complex structures!  Moreover, the transformation (\ref{U(1)}) is not just the usual diffeomorphism on $\chi^i$, but it includes an additional chiral rotation.  In mathematical language, one says that $\chi^i$ is a section of a U(1) bundle over the hyper-K\"ahler manifold.

The action is determined by supersymmetry and the fermion equations of motion (\ref{AdS Dirac}).  It is given by (\ref{action}), where
the potential ${\cal V}$ is now
\begin{equation}
{\cal V} =  g_{ij^*}X^i\bar{X}^{j^*} + kD(A,A^*),
\end{equation}
and
\begin{equation}
4ig_{ij^*}\bar{X}^{j^*} = \frac{\partial}{\partial A^i} D(A,A^*) .
\end{equation}
Equation (\ref{AdSKillingCondition}) implies that $D_i$ is integrable.  The total action is invariant under supersymmetry when the holomorphic Killing vector $X^i$ satisfies the inhomogeneous tri-holomorphic Killing condition \cite{BX}.

For a given hyper-K\"ahler manifold, one would like to solve (\ref{Modifiedtri-holo}) and (\ref{AdSKillingCondition}) to find all possible Killing vectors $X^i$.  The task is simple when the manifold admits a holomorphic homothetic Killing vector $Y^i$ such that
\begin{equation}\label{Ytri-holo}
\nabla_jY^i=\delta^i{}_j.
\end{equation}
The $X^i$ can then written as
\begin{equation}\label{Y+Z}
X^i=Z^i-i\frac{3k}{2}Y^i,
\end{equation}
where $Z$ is a tri-holomorphic Killing vector that satisfies the usual  tri-holomorphic condition (\ref{tri-holo condition}). Such manifolds are known as hyper-K\"ahler cones or Swann spaces \cite{dewit}.  Note that in AdS$_5$, there is a nonvanishing potential even when $Z^i=0$.

\section{Superconformal Symmetry on the Brane}

The supergroup of AdS$_5$ isometries, SU(2,2$|$1), is also the superconformal group in four flat spacetime dimensions.  It is important, therefore, to look at the hypermultiplet coupling from a superconformal point of view.  The analysis is helped by the fact that with our metric, the slices $z=c$ (for constant $c$) foliate AdS$_5$ into a set of 3-branes, each  flat and Minkowski.  The bulk fields $A^i(x,z)$ and $\chi^i(x,z)$, for fixed $z=c$, are four-dimensional brane fields.  In this section, we will see how the on-shell AdS$_5$ supersymmetry transformations give rise to off-shell four-dimensional superconformal transformations for the brane fields after suitable boundary conditions are imposed.

In two-component notation, the AdS$_5$ supersymmetry transformations take the following form:
\begin{eqnarray}\label{adstrans2}
\frac{1}{\sqrt{2}}\delta A^i&=&e^{-\frac{1}{2}kz}\epsilon\chi^i+ike^{-\frac{1}{2}kz}x^m\delta^a_m\bar{\eta}\bar{\sigma}_a\chi^i+e^{\frac{1}{2}kz}\Omega^i{}_{j^*}\bar{\eta}\bar{\chi}^{j^*}\nn \\
e^{-\frac{1}{2}kz}\frac{1}{\sqrt{2}}\delta \chi^i&=&i\delta^m_a\sigma^a\bar{\epsilon}\partial_mA^i+kx^n\delta^{a}_n\eta^{bm}\sigma_b\bar{\sigma}_a\eta\partial_mA^i-\eta(\partial_5A^i+iX^i)\nn\\
&&\ -\ ie^{kz}\Omega^i_{\ j^*}\delta^m_a\sigma^a\bar{\eta}\partial_mA^{*j^*}-\Gamma^i_{jk}\Omega^j_{\ p^*}\bar{\eta}\bar{\chi}^{p^*}\chi^k\nn\\
&&\ -\ \epsilon\left[\Omega^i{}_{j^*}\left(\partial_5A^{*j^*}+i\bar{X}^{j^*}\right)-\frac{1}{2}\Gamma^i_{jk}\chi^j\chi^k\right]\nn\\
&&\ +\ ik x^m\delta^a_m\sigma_a\bar{\eta}\left[\Omega^i{}_{j^*}\left(\partial_5A^{*j^*}+i\bar{X}^{j^*}\right)-\frac{1}{2}\Gamma^i_{jk}\chi^j\chi^k\right].
\end{eqnarray}
In this expression, the $A^i$ are the coordinates of a hyper-K\"ahler manifold, and $X^i$ is a holomorphic Killing vector that satisfies the inhomogeneous tri-holomorphic Killing condition (\ref{Modifiedtri-holo}).  On the brane at $z=c$, we seek an ${\cal N}=1$ theory in which the scalar fields are coordinates of a K\"ahler manifold, not necessarily hyper-K\"ahler.  We find such a theory by imposing boundary conditions on the brane that
\begin{enumerate}
\item
eliminate half of the fermionic fields, and
\item
transform half of the bulk bosons into auxiliary fields.
\end{enumerate}
In this way we transform an on-shell bulk theory into an off-shell theory on the brane.

The boundary conditions we impose are as follows:
\begin{eqnarray}\label{bc}
0 &=& \Omega^I_{\ j^*}\bar{\chi}^{j^*}\big|_{z=c} \nn\\
W^I &=& (-i\partial_5 A^I + X^I)\big|_{z=c} \nn\\
F^I &=&  e^{-kz}\left.\left(-\Omega^I{}_{j^*}\partial_5 A^{j^*}-i\Omega^{I}{}_{j^*}\overline{X}^{j^*}+\frac{1}{2}\Gamma^{I}_{jk}\chi^j\chi^k\right)\right|_{z=c}
\end{eqnarray}
where $W^I$ is a holomorphic vector, $I = 1,...,n$, and the lower-case indices run from 1 to $2n$.  Closure of the bulk algebra imposes the additional constraints:
\begin{eqnarray}\label{2ndbc}
0 &=& \left.\left(i\Omega^I{}_{j^*} \delta_a^m\sigma^a\partial_mA^{*j^*}+e^{-kz}\Gamma^{I}_{jk}\Omega^j{}_{l^*}\chi^k\bar{\chi}^{l^*}\right) \right|_{z=c}\nn\\
0&=&\left[ie^{kz}\Gamma^{I}_{pq}\Omega^p{}_{r^*}\delta^{m}_a\bar{\sigma}^a\chi^q\partial_mA^{r^*}+\Gamma^I_{pq}\Omega^p{}_{r^*}\Omega^q{}_{j^*}
\left(\partial_5A^{j^*}+i\overline{X}^{j^*}\right)\bar{\chi}^{r^*}\right.\nn\\
&&\left.\left.+\frac{1}{2}\left(\partial_k\Gamma^{I}_{pq}-\Gamma^{I}_{pt}\Gamma^{t}_{kq}-\Gamma^{I}_{qt}\Gamma^{t}_{kp}\right)
(\chi^k\chi^q)\Omega^p{}_{r^*}\bar{\chi}^{r^*}\right]\right|_{z=c}\nn\\
W^{I}_{J}\chi^J&=&\left.\left(-i\partial_5\chi^I+i\frac{k}{2}\chi^I+X^{I}_{j}\chi^j\right)\right|_{z=c}
\end{eqnarray}
The mixed Dirichlet-Neuman conditions eliminate half the fermionic fields, while leaving the scalar fields unconstrained.  The third condition in (\ref{2ndbc}) defines the superconformal $R$-transformation on the field $\chi^I$.  (These constraints can also be derived from the ${\cal N} = 1$ superspace transformations \cite{BX}.)

With these boundary conditions, we see immediately that the scalar field transformation becomes
\begin{equation}\label{SUSYA}
\frac{1}{\sqrt{2}}\delta A^I\big|_{z=c}\ =\ e^{-\frac{1}{2}kz}\epsilon\chi^I+ike^{-\frac{1}{2}kz}x^m\delta^a_m\bar{\eta}\bar{\sigma}_a\chi^I,
\end{equation}
as required for a superconformal theory.  The fermion transformation is more complicated.  Applying the boundary conditions (\ref{bc}) and (\ref{2ndbc}) to (\ref{adstrans2}), we find
\begin{eqnarray}
e^{-\frac{1}{2}kz} \frac{1}{\sqrt{2}}\delta \chi^I\big|_{z=c}&=&i\delta^m_a\sigma^a\bar{\epsilon}\partial_mA^I+k x^n\delta^{a}_n\eta^{bm}\sigma_b\bar{\sigma}_a\eta\partial_mA^I-i \eta W^I \nn\\
&&\ +\ \epsilon F^I - ik x^m\delta^a_m\sigma_a\bar{\eta} F^I.
\end{eqnarray}
The auxiliary field $F^I$ is defined in (\ref{bc}).  Its transformation can be computed with the help of (\ref{2ndbc}) and the five-dimensional fermionic equations of motion.  We find
\begin{equation}
e^{\frac{1}{2}kz} \frac{1}{\sqrt{2}}\delta F^I\big|_{z=c}\ =\ i \delta^{m}_a\bar{\epsilon}\bar{\sigma}^{a}\partial_{m}\chi^{I}-k x^{n}\delta_{n}^{b}\delta^{m}_{a}\eta\sigma_b\bar{\sigma}^a\partial_m\chi^I -2k \eta\chi^I+2i W^{I}_{J}\eta\chi^J .
\end{equation}

The supersymmetry transformations on the brane are precisely those of $n$ off-shell superconformal chiral multiplets \cite{Sezgin} $(A^I, e^{-\frac{1}{2}kz}\chi^I, F^I)$.  The $n$ on-shell AdS$_5$ hypermultiplets reduce to $n$ off-shell four-dimensional superconformal chiral multiplets.  Although $X^i$ is Killing, and satisfies the inhomogeneous tri-holomorphic condition (\ref{Modifiedtri-holo}), the vector $W^I$ is not similarly restricted.  Note that all the $\Omega$ and $\Gamma$ are absorbed in the $F^I$, so the hyper-K\"ahler structure of the bulk hypermultiplets disappears in the off-shell ${\cal N}=1$ transformations on the brane.

\section{Summary}

In this paper we used component fields to demonstrate that the supersymmetric nonlinear sigma model in AdS$_5$ is described by hyper-K\"ahler geometry.  We have seen that the target-space manifold must admit a holomorphic Killing vector $X^i$ that satisfies the inhomogeneous tri-holomorphic condition (\ref{Modifiedtri-holo}).  Our work confirms and extends results that were previously found in superspace \cite{BX}.

The AdS$_5$ hypermultiplet enjoys a close connection to the four-dimensional superconformal chiral multiplet.  In this paper we also proposed a set of boundary conditions that reduce the {\it on-shell} AdS$_5$ hypermultiplet transformations to {\it off-shell} superconformal transformations on ${\cal N}=1$ chiral multiplets.

We would like to thank Chi Xiong for useful discussions.  This work was supported in part by the U.S.\ National Science Foundation, grant NSF-PHY-0910467.

\end{document}